\documentclass[11pt]{article}
\begin{document}
\thispagestyle{empty}
\begin{flushright}
MPI-PhT/96-23\\
UNIGRAZ-UTP-03-04-96\\
hep-lat/9604002
\end{flushright}
\vskip 1cm
\begin{center}
{\huge{Lattice Regularization of the Chiral Schwinger Model }} 
\vskip10mm
\centerline{ {\bf
Christof Gattringer${}^*$ }}
\vskip 5mm
\centerline{Max-Planck-Institut f\"{u}r
Physik (Werner-Heisenberg-Institut)}
\centerline{F\"ohringer Ring 6, 80805 Munich, Germany}
\vskip 2mm
\centerline{and}
\vskip 2mm
\centerline{Institut f\"{u}r Theoretische Physik der Universit\"at Graz} 
\centerline{Universit\"atsplatz 5, 8010 Graz, Austria}
\vskip35mm
\end{center}
\begin{abstract}
\noindent
We analyze the chiral Schwinger model on an infinite lattice 
using the continuum definition of the fermion determinant and 
a linear interpolation of the lattice gauge fields. For non-compact 
and Wilson formulation of the gauge field action it is proven that 
the effective lattice model is Osterwalder-Schrader 
positive, which is a sufficient condition for the reconstruction 
of a physical Hilbert space from the model defined on a Euclidean lattice. 
For the non-compact model we furthermore establish the existence of 
critical points where the corresponding continuum theory can be reconstructed.
We show that  the continuum limit for the two-point functions of field 
strength and chiral densities can be controlled analytically. The article ends
with some remarks on fermionic observables.
\end{abstract}
\vskip20mm
\bigskip \nopagebreak \begin{flushleft} \rule{2 in}{0.03cm}
\\ {\footnotesize \ 
${}^*$ e-mail: chg@mppmu.mpg.de}
\end{flushleft}
\newpage
%
%
\section{Introduction}
It has been known since the first days of lattice field theory 
\cite{wilson} that the regularization of chiral field theories
on a lattice is a notoriously difficult problem (see \cite{shamir}
for a recent review). The celebrated Nielsen-Ninomya theorem 
\cite{nini} states that under reasonable assumptions (e.g locality of
the lattice action) the number 
of left-handed fermions equals the number of right-handed fermions
in a fermionic lattice theory. Thus it seems impossible to regularize
chiral field theories as e.g. the electroweak sector of the standard 
model on the lattice.

In \cite{thooft} 't Hooft renewed the interest in an old, 
alternative approach \cite{flume}-\cite{stara}. The idea is to put only
the gauge field on the lattice, interpolate the gauge field to 
the interior of the lattice cells and couple the interpolated 
gauge field to continuum fermions. This hybrid approach\footnote{
The author partly ows the term "hybrid approach" to R.~L.~Stuller who 
christened his approach "Hybrid Quantization".} has many 
advantages. The number of degrees of freedom for the gauge fields
remains finite (or countably infinite) and one can construct a 
measure for the gauge fields, which is (to the knowledge of the author)
not possible directly in the continuum except for the gauge group U(1) 
(Gaussian measures). The Nielsen-Ninomya theorem is circumvented, 
since the continuum fermions do not have a local action from a lattice 
point of view. The hard part of the hybrid approach is to find a 
proper and explicit definition of the continuum fermion determinant 
in the interpolated background
field. In \cite{thooft} it is argued that for the U(1) 
gauge theory in 4 dimensions with 
vectorlike coupling to the fermions the continuum determinant in the 
interpolated background field can be given a meaning, and that all 
chiral symmetries are kept intact. This is an important result,
but unfortunately the construction is not explicit, and it is not
possible to explore details analytically. However it has to be remarked 
that similar, more explicit constructions using a finer grid for the 
fermions are being studied \cite{hernan}. 
 
When proposing such a conceptionally new approach, one has to analyze its
fundamental properties that allow to get back to a continuum theory
in Minkowski space. In particular one is interested in proving
Osterwalder-Schrader (OS) positivity \cite{osterschra}
which allows the reconstruction of the physical Hilbert space and 
Hamiltonian from a theory with Euclidean metric. Furthermore one has to
establish the
existence of critical points with diverging correlation length, where the 
continuum limit can be performed. Analyzing such properties in four
dimensional theories is not possible at the moment, in particular since 
explicit results for the fermion determinant in the interpolated 
background field are not available yet. However it is already desirable
to see the hybrid approach work in simple models which can be controlled 
analytically. In \cite{latsch} it was shown that for the hybrid
approach to the vectorlike Schwinger model a critical point exists 
where the continuum limit can be performed. It was furthermore
established that the chiral symmetries are kept intact and that no 
doubling occurs. A proof of OS positivity for the vectorlike model was 
given in \cite{stara}.

The next logical step is to test the hybrid approach in a chiral model. 
A perfect candidate is the chiral Schwinger model first analyzed in 
\cite{jara} - \cite{grr}. The chiral Schwinger model in the continuum is under 
good analytic control, in particular the fermion determinant is known
\cite{jara}. Results from the hybrid approach can be compared 
to their continuum counterparts explicitly. It has to be remarked that 
\cite{gock} attempted to  set up the hybrid approach in this model long 
before the work of 't Hooft \cite{thooft}. 

In this paper we apply the hybrid approach to the chiral Schwinger 
model. We use an interpolation scheme \cite{flume,kronfeldal,latsch} 
which was
already successfully implemented in the vectorlike model 
\cite{latsch, stara}. The Jackiw-Rajaraman determinant \cite{jara}
is used to construct the effective lattice gauge theory (Section 2).
In Section 3 we prove that the resulting effective lattice gauge 
model is OS positive for both the non-compact and the Wilson
formulation of the gauge field action. In Section 4 we establish 
the existence of critical points and perform the continuum limit
for the two point function of the field strength. Section 5 is devoted 
to the discussion of fermionic observables. Section 6 contains 
concluding remarks.  
%
%
\section{Interpolation and the effective lattice action}
\setcounter{equation}{0}
The lattice under consideration is
$\mbox{Z\hspace{-1.35mm}Z}^2$, i.e. the lattice spacing is set to one.
Functions defined in the continuum can be identified by their 
arguments $x,y \in \mbox{I\hspace{-0.7mm}R}^2$, 
while lattice quantities have arguments
$n,m \in \mbox{Z\hspace{-1.35mm}Z}^2$.
We consider two types of actions for the U(1)-gauge fields.
The non-compact action is given by
\begin{equation}
S_{nc} \; := \; \frac{1}{4} \sum_{n \in Z\mbox{\hspace{-1.35mm}}Z^2}
\; F_{ \mu \nu}(n) F_{ \mu \nu}(n) \; = \; 
\frac{1}{2} \sum_{n \in Z\mbox{\hspace{-1.35mm}}Z^2}
\; F_{12}(n)^2 \; \;  ,
\end{equation}
with
\begin{equation}
F_{\mu \nu}(n) \;  := \;  \Big( A_\nu(n + \hat{e}_\mu) -
A_\nu(n) - A_\mu(n + \hat{e}_\nu) + 
A_\mu(n) \Big) \; .
\end{equation}
$A_\mu(n)$ may assume values in $(-\infty, + \infty)$ for all 
$n \in \mbox{Z\hspace{-1.35mm}Z}^2$ 
and $\mu = 1,2$. $\hat{e}_\mu$ is the unit
vector in $\mu$-direction. 
We also consider the Wilson formulation of the gauge field action
\begin{equation}
S_{w} \; := \; \frac{1}{ e^2} \sum_{n \in Z\hspace{-1.35mm}Z^2}
\; \left[ 1 \; - \; \mbox{Re} \left( U_1(n) U_2(n\!+\!\hat{e}_1)
\overline{ U_1(n\!+\!\hat{e}_2)} 
\overline{ U_2(n)} \right) \right] 
\; = \; \frac{1}{ e^2} \sum_{n \in Z\hspace{-1.35mm}Z^2}
\; \left[ 1 \; - \; \mbox{Re} \; e^{i e F_{12}(n)} \right] \;. 
\end{equation}
In the last step the gauge transporters were expressed as
\begin{equation}
U_\mu(n) \; := \; \exp\left( ie \; A_\mu(n) \right) \; \; \; 
\mbox{with} \; \; \; \; e A_\mu(n) \; \in \; [ 0, 2\pi ) \; .
\end{equation}
Although we use the same symbol, it has to be kept in mind that the 
$A_\mu({n})$ are restricted to the principal branch $[ 0, 2\pi/e)$ 
in the case of the Wilson action. 

We interpolate the
fields as follows \cite{flume,kronfeldal,latsch}
\begin{equation}
A_1^{int}(x) \; := \; A_1(n) \;  [1-t_2]  \;  + \;  
A_1(n+\hat{e}_2) t_2 \; \; \; \; , \; \; \; \; 
A_2^{int}(x) \; := \; A_2(n) \;  [1-t_1]  \;  + \;  
A_2(n+\hat{e}_1) t_1 \; \; ,
\end{equation}
for $x = n + t $ and $t_1,t_2 \in (0,1]$. This type of interpolation was
already used for the treatment of the vectorlike Schwinger model on the lattice
\cite{flume,latsch}. In \cite{latsch} it was shown that the interpolation
(2.5) respects the condition of gauge equivariance (i.e. transforming a lattice
gauge transformation to a continuum gauge transformation for the interpolated
fields).

For later use we quote the (continuum) Fourier transform of the interpolated
fields
\[
\widetilde{A}_1^{int}(p) \; = 
\; \int_{-\infty}^\infty d^2x A_1(x)e^{-ipx} \; =: \; 
\widehat{A}_1(p) \; 
\frac{1-e^{-ip_1}}{ip_1} \frac{2-2\cos(p_2)}{p_2^2} \; ,
\]
\begin{equation}
\widetilde{A}_2^{int}(p) \; = \; \int_{-\infty}^\infty 
d^2x A_2(x)e^{-ipx} \; =: \; 
\widehat{A}_2(p) \; \frac{1-e^{-ip_2}}{ip_2} \frac{2-2\cos(p_1)}{p_1^2} \; ,
\end{equation}
where we introduced the lattice Fourier transform
\begin{equation}
\widehat{A}_\mu(p) \; := \; \sum_{n \in Z\mbox{\hspace{-1.35mm}}Z^2}
e^{-ipn} A_\mu(n) \; .
\end{equation}
The continuum Fourier transform $\widetilde{A}_\mu^{int}$
of the interpolated lattice fields comes
out as the Fourier transform on the lattice
$\widehat{A}_\mu$ multiplied by some function 
which depends on the details of the interpolation and explicitly
shows the ultraviolet regulator (inverse powers of $p_\mu$) 
which is introduced by the lattice.

The first step in the hybrid approach is to give a meaning to the 
determinant in the continuum. The Euclidean continuum action for the fermions
is given by 
\begin{equation}
S_F \; := \; \int_{-\infty}^\infty 
d^2x \; \overline{\chi}(x) \; \Big[
\partial_1 - i\partial_2 -ie\Big(A_1(x) - iA_2(x)\Big) \Big] \; 
\chi(x) \; ,
\end{equation}
where $\overline{\chi}, \chi$ are left handed Weyl fermions. It has to 
be remarked, that often (see e.g. \cite{jara})) 
the chiral Schwinger model is
defined using Dirac spinors, where only one chiral component couples to
the gauge field. Of course the results are the same, since the other
component decouples. The well known result for the regularized fermion 
determinant is given by \cite{jara} 
(see e.g. \cite{linha} for the translation to the
Euclidean version quoted here)
\begin{equation}
\mbox{det}_{reg}\left[ \partial_1 \! - \! i\partial_2 \! -ie \! 
\left(A_1 \! -  \!
iA_2 \right) \right] =  \exp \! \left(\!-\frac{1}{2} g^2 \! \!
\int_{-\infty}^\infty \!
\frac{d^2p}{(2\pi)^2}\frac{1}{p^2} \widetilde{A}_\mu(-p)
\widetilde{M}_{\mu \nu}(p) \widetilde{A}_\nu(p)\!\right) =:  
\exp\left(-W[A]\right) ,
\end{equation}
where
\begin{equation}
\widetilde{M}_{\mu \nu}(p) \; = \; 
(a+1) \delta_{\mu \nu} p^2 \; - \; 2 p_\mu p_\nu \; + \; 
i\left[ p_\mu \varepsilon_{\rho \nu} p_\rho + 
p_\rho \varepsilon_{\rho \mu} p_\nu \right] \; .
\end{equation}
$g^2$ is defined as $g^2 := e^2/4\pi$. The Jackiw-Rajaraman parameter 
$a$ parametrizes 
an ambiguity in the regularization of a short distance singularity.
An explicit computation of the determinant directly in Euclidean space, 
following 
\cite{alvarez} can be found in \cite{gock}. There the regularization is 
performed such that $a = 1$. The full freedom in the regularization of the 
short distance singularity is e.g. analyzed in \cite{banerjee}. 

The effective action $W[A]$ defined through (2.9), (2.10) is complex
and cannot be made gauge invariant by adjusting $a$. It has to
be remarked, that for $a = 1$ the kernel of the quadratic form in 
the gauge fields $W[A]$ cannot be inverted, and the propagator for the
gauge fields does not exist then. For $a < 1$, the model contains ghosts 
\cite{jara}.

The next step in the hybrid approach for formulating the chiral model on 
the lattice is to plug the interpolated gauge fields (2.6) into
the expression (2.9), (2.10) for the continuum determinant giving rise to
\begin{equation}
W^{lat}[A] \; := \; W[A^{int}] \; = \;
\frac{1}{2} g^2 \int_{-\infty}^\infty
\frac{d^2p}{(2\pi)^2}
\frac{2-2\cos(p_1)}{p_1^2} \frac{2-2\cos(p_2)}{p_2^2}
\frac{1}{p^2} \; \widehat{A}_\mu(-p)
\widehat{M}_{\mu \nu}(p) \widehat{A}_\nu(p) \; ,
\end{equation}
where the kernel $\widehat{M}_{\mu \nu}$ for the lattice fields
$\widehat{A}_\mu$ is given by
\begin{eqnarray}
\widehat{M}_{11}(p) \; & = & \; 
\left[ (a+1)p^2 - 2p_1^2 - i2p_1p_2 \right]\frac{2-2\cos(p_2)}{p_2^2} \; ,
\nonumber \\
\widehat{M}_{12}(p) \; & = & \; 
\left[ - 2p_1p_2 + i(p_1^2 - p_2^2) \right]
\frac{(e^{ip_1}-1)(e^{-ip_2}-1)}{p_1p_2} \; ,
\nonumber \\
\widehat{M}_{21}(p) \; & = & \; 
\left[ - 2p_1p_2 + i(p_1^2 - p_2^2) \right]
\frac{(e^{-ip_1}-1)(e^{ip_2}-1)}{p_1p_2} \; ,
\nonumber \\
\widehat{M}_{22}(p) \; & = & \; 
\left[ (a+1)p^2 - 2p_2^2 + i2p_1p_2 \right]\frac{2-2\cos(p_1)}{p_1^2} \; .
\end{eqnarray}
For later use we quote the following symmetry properties of 
$\widehat{M}_{\mu \nu}$
\begin{equation}
\widehat{M}_{\mu \mu}(-p_1, p_2) \; = \; 
\widehat{M}_{\mu \mu}(p_1, -p_2) \; = \; 
\overline{\widehat{M}_{\mu \mu}(p_1, p_2)} 
\; \; \; \; , \; \; \; \; \mu = 1,2 \; ,
\end{equation}
\begin{equation}
\widehat{M}_{12}(-p_1, -p_2) \; = \; 
\widehat{M}_{21}(p_1, p_2) \; , 
\end{equation}
\begin{equation}
-e^{-ip_2} \; \overline{\widehat{M}_{21}(p_1, -p_2)} \; = \; 
\widehat{M}_{12}(p_1, p_2) \; \; \; \; , \; \; \; \;
-e^{ip_2} \; \overline{\widehat{M}_{12}(p_1, -p_2)} \; = \; 
\widehat{M}_{21}(p_1, p_2) \; .
\end{equation}
Adding the contribution $W^{int}$ from the fermion determinant to the 
lattice action for the gauge fields $S_{nc}$, $S_{w}$ respectively 
gives the effective lattice action for the chiral Schwinger model in 
non-compact and Wilson formulation
\begin{equation}
S^{eff}_{nc} \; := \; S_{nc} + W^{lat} \; \; \; \; , \; \; \; \;
S^{eff}_{w} \; := \; S_{w} + W^{lat} \; .
\end{equation}

%
%

\section{Proof of OS positivity}
\setcounter{equation}{0}
Osterwalder and Schrader \cite{osterschra} 
developed a mathematical procedure that 
allows the reconstruction of the Hamiltonian and the physical 
Hilbert space from a continuum field theory defined in Euclidean space.
The main condition is Osterwalder-Schrader (OS) positivity (see below).
For the Wilson formulation of lattice gauge theory,  
OS positivity was shown to hold by Osterwalder and Seiler 
\cite{oster}. 
For the hybrid approach a new proof has to be given. 

For this purpose, we decompose the field algebra into three disjoint sets
${\cal A}^+, {\cal A}^0$ and ${\cal A}^-$, defined as
\begin{eqnarray}
{\cal A}^+ \; & := & \; \left\{ A_1(n_1,n_2), A_2(n_1,m_2) \Big|
n_1 \in \mbox{Z\hspace{-1.35mm}Z}, n_2 > 1, m_2 > 1 \right\} \; ,
\nonumber \\
\!{\cal A}^0 \; & := & \; \left\{ A_1(n_1,1), A_1(n_1,0),
A_2(n_1,1), A_2(n_1,0), A_2(n_1,-1) \Big|
n_1 \in \mbox{Z\hspace{-1.35mm}Z} \right\} \; , \nonumber \\
{\cal A}^- \; & := & \; \left\{ A_1(n_1,n_2), A_2(n_1,m_2) \Big|
n_1 \in \mbox{Z\hspace{-1.35mm}Z}, n_2 < 0, m_2 < -1 \right\} \; .
\end{eqnarray}
We furthermore define the antilinear time reflection operator $\Theta$
acting on the gauge fields as follows
\begin{equation}
\Theta A_1(n_1,n_2) \; := \; A_1(n_1,-n_2+1) \; \; \; , \; \; \;
\Theta A_2(n_1,n_2) \; := \; -A_2(n_1,-n_2) \; .
\end{equation}
It is easy to check that $\Theta$ maps ${\cal A}^+$ onto ${\cal A}^-$
and ${\cal A}^0$ onto itself. 
The condition of OS positivity is now defined as 
\begin{equation}
\Big\langle P[{\cal A}^+] \; \Theta P[{\cal A}^+] \Big\rangle
\; \geq \; 0 \; \; \; \; , \; \; \; \; \forall \; P[{\cal A}^+] \; ,
\end{equation}
where $P[{\cal A}^+]$ denotes an arbitrary function depending 
only on the field variables in ${\cal A}^+$. Following 
\cite{oster, froh} we show that both effective actions (2.16) can be 
decomposed as follows
\begin{equation}
-S^{eff} \; = \; -S^+[{\cal A}^+, {\cal A}^0] \; - \; \Theta
S^+[{\cal A}^+, {\cal A}^0] \; + \; 
\int d\mu[\lambda] \; C_\lambda[{\cal A}^+] \; 
\Theta C_\lambda[{\cal A}^+] \; ,
\end{equation}
where $d\mu[\lambda]$ denotes some positive measure, 
and $S^+[{\cal A}^+, {\cal A}^0]$
depends only on the field variables in ${\cal A}^+ \cup
{\cal A}^0$ and $C_\lambda[{\cal A}^+]$ only on the variables in ${\cal A}^+$.
The decomposition (3.4) is a sufficient condition for (3.3) 
to hold, since by expanding the
exponential of the last term in (3.4) one obtains (use the antilinearity of
$\Theta$)
\begin{equation}
\Big\langle P[{\cal A}^+] \; \Theta P[{\cal A}^+] \Big\rangle  = 
\frac{1}{Z} \int \! D[{\cal A}^0]\!\sum_{j=0}^\infty\frac{1}{j!} \! 
\int \! \prod_{l=1}^j
d\mu[\lambda_l] 
\left| \int \! D[{\cal A}^+] e^{-S^+[{\cal A}^+, {\cal A}^0]} 
P[{\cal A}^+] \prod_{i=1}^j C_{\lambda_i}[{\cal A}^+]
\right|^2 \; \geq \; 0 .
\end{equation}

For the two expressions $S_{nc}, S_{w}$ of the gauge field part of 
the action, it is well known, that a decompositions of the type 
(3.4) exists \cite{oster, stara}.
Using $\Theta F_{12}(n_1,n_2) = - F_{12}(n_1,-n_2)$, which follows from 
(3.2), one finds
\begin{equation}
-S_{nc} \; = \; -S^+_{nc} \; - \; \Theta S^+_{nc} \; \; \; \; , 
\; \; \; \; -S_{w} \; = \; -S^+_{w} \; - \; \Theta S^+_{w} \; ,
\end{equation}
where 
\begin{equation}
S^+_{nc} \; := \; \frac{1}{2} \sum_{n_1 \in Z\hspace{-1.35mm}Z, n_2 > 0}
F_{12}(n_1,n_2)^2 \; + \; \frac{1}{4} \sum_{n_1 \in
Z\hspace{-1.35mm}Z} F_{12}(n_1,0)^2 \; ,
\end{equation}
and
\begin{equation}
S^+_{w} \; := \; \frac{1}{e^2} \sum_{n_1 \in Z\hspace{-1.35mm}Z, n_2 > 0}
\Big[1 - \mbox{Re} \; e^{ie F_{12}(n_1,n_2)} \Big] \; + \; 
\frac{1}{2e^2} \sum_{n_1 \in Z\hspace{-1.35mm}Z}
\Big[1 - \mbox{Re} \; e^{ie F_{12}(n_1,0)} \Big] \; .
\end{equation}
Thus there is only the contribution $W^{lat}[A]$ from the fermion determinant 
left to analyze. Using (3.2), and the antilinearity of $\Theta$,
we decompose the Fourier transform (2.7) of the gauge field components as 
\begin{eqnarray}
\widehat{A}_1(p_1,p_2) \; & = & \; \widehat{A}^+_1(p_1,p_2) \; + \; 
\widehat{A}^0_1(p_1,p_2) \; + \; 
e^{-ip_2} \Theta \widehat{A}^+_1(-p_1,p_2) \; , 
\\
\widehat{A}_2(p_1,p_2) \; & = & \; \widehat{A}^+_2(p_1,p_2) \; + \; 
\widehat{A}^0_2(p_1,p_2) \; - \; 
\Theta \widehat{A}^+_2(-p_1,p_2) \; , 
\end{eqnarray}
where we defined
\begin{equation}
\widehat{A}^+_\mu(p) \; \; := \; \sum_{n_1 \in Z\hspace{-1.35mm}Z, 
n_2 > 1} e^{-ipn} A_\mu(n) \; , \; \; \; \mu = 1,2 \; ,
\end{equation}
and
\begin{eqnarray}
\widehat{A}^0_1(p_1,p_2) \; & := & \; 
\sum_{n_1 \in Z\hspace{-1.35mm}Z} e^{-ip_1n_1} 
\Big[ e^{-ip_2} A_1(n_1,1) + A_1(n_1,0)  \Big] \; , \\
\widehat{A}^0_2(p_1,p_2) \; & := & \; 
\sum_{n_1 \in Z\hspace{-1.35mm}Z} e^{-ip_1n_1} 
\Big[ e^{-ip_2} A_2(n_1,1) + A_1(n_1,0) + e^{ip_2} A_2(n_1,-1) \Big]
\; .
\end{eqnarray}
For later use we note the reflection properties of the 
$\widehat{A}^0_\mu$ which follow from (3.2) and the antilinearity
of $\Theta$
\begin{equation}
\Theta \widehat{A}^0_1(p_1,p_2) \; = \; 
e^{ip_2} \widehat{A}^0_1(-p_1,p_2) \; \; \; \; , \; \; \; \; 
\Theta \widehat{A}^0_2(p_1,p_2) \; = \; 
- \widehat{A}^0_2(-p_1,p_2) \; .
\end{equation}
Inserting (3.9) and (3.10) into (2.11) gives, after writing explicitly
the sum over $\mu$ and $\nu$, altogether 36 terms. Some of them are 
equal, some are the image of others under the reflection $\Theta$. Using
(3.14), the properties (2.13)-(2.15) of $\widehat{M}_{\mu \nu}$,  
the antilinearity of $\Theta$ and coordinate transformations of the type
$p_\mu \rightarrow -p_\mu$ in the integrals over the momentum space, one 
obtains after a lengthy but straightforward computation
\begin{equation}
-W^{lat}[A] \; = \; -W[A]^+ \; - \; \Theta W[A]^+ \; + \; W^{mix}[A] \; ,
\end{equation}
with
\[
W[A]^+ \; := \; \frac{g^2}{4} \int_{-\infty}^\infty 
\frac{d^2p}{(2\pi)^2} \; \frac{1}{p^2} 
\frac{2-2\cos(p_1)}{p_1^2} \frac{2-2\cos(p_2)}{p_2^2} \]
\begin{equation}
\times \Big[ 
2 \widehat{A}_\mu^+ (-p) \widehat{M}_{\mu \nu}(p) 
\widehat{A}_\nu^+ (p) \; + \; 
4 \widehat{A}_\mu^0 (-p) \widehat{M}_{\mu \nu}(p) 
\widehat{A}_\nu^+ (p) \; + \;
\widehat{A}_\mu^0 (-p) \widehat{M}_{\mu \nu}(p) 
\widehat{A}_\nu^0 (p) \Big] \; .
\end{equation}
Thus (compare (3.4) and (3.15)) there is only left to show, that 
$W^{mix}[A]$ has the form of the last term in (3.4). $W^{mix}[A]$
is given by
\begin{equation}
W^{mix}[A] \; := \; g^2 \int_{-\infty}^\infty \frac{d^2p}{(2\pi)^2} \;
\frac{1}{p^2}
\frac{2-2\cos(p_1)}{p_1^2} \frac{2-2\cos(p_2)}{p_2^2} \Big[ 
\Theta \widehat{A}_\mu^+ (p_1,-p_2) \Big] 
\widehat{R}_{\mu \nu}(p_1,p_2) \widehat{A}_\nu^+ (p_1,p_2) \; ,
\end{equation}
where
\begin{eqnarray}
\widehat{R}_{11}(p) \; & := & \; \Big[ -(a+1)p^2 + 2(p_1^2 + ip_1 p_2) \Big] \;
\frac{2-2\cos(p_2)}{p_2^2} e^{ip_2} \; , \nonumber \\
\widehat{R}_{12}(p) \; & := & \; \Big[ 2p_1 p_2 -i (p_1^2 - p_2^2) \Big] \;
\frac{(e^{ip_1} - 1 )( e^{-ip_2} - 1)}{p_1 p_2} e^{ip_2} \; , \nonumber \\
\widehat{R}_{21}(p) \; & := & \; \Big[ - 2p_1 p_2 + i (p_1^2 - p_2^2) \Big] \;
\frac{(e^{-ip_1} - 1 )( e^{ip_2} - 1)}{p_1 p_2} \; , \nonumber \\
\widehat{R}_{22}(p) \; & := & \; \Big[ (a+1)p^2 - 2(p_2^2 - ip_1 p_2) \Big] \;
\frac{2-2\cos(p_1)}{p_1^2} \; .
\end{eqnarray}
Inserting (3.11) into (3.17) one obtains (use (3.2), and
the antilinearity of $\Theta$)
\begin{equation}
W^{mix}[A] \; = \; \sum_{n_1,m_1 \in Z\hspace{-1.35mm}Z} 
\sum_{n_2,m_2>1}
\Big[ \Theta A_\mu(m_1,m_2) \Big] A_\nu(n_1,n_2) 
P_{\mu \nu}(m_1-n_1,m_2+n_2) \; ,
\end{equation}
where
\begin{equation}
P_{\mu \nu}(k_1,k_2) \; := \; 
g^2 \int_{-\infty}^\infty \frac{dp_1}{(2\pi)^2} 
\frac{2-2\cos(p_1)}{p_1^2} e^{ip_1k_1} 
\int_{-\infty}^\infty \! dp_2 \frac{2-2\cos(p_2)}{p_2^2} e^{-ip_2k_2}
\frac{1}{p_1^2 + p_2^2} \widehat{R}_{\mu \nu}(p_1,p_2) \; .
\end{equation}
The $p_2$ integration can be carried out using the residue theorem. We close
the contour in the lower complex $p_2$-half plane. The  
factor $e^{-ip_2k_2}$ dominates the exponentials from the cosine and the 
exponents in $\widehat{R}_{\mu \nu}(p_1,p_2)$, since $k_2 = m_2 + n_2 \geq 4$
as can be seen from the sum in (3.19). 
The only pole in the lower $p_2$-half plane is at 
$p_2 = -i |p_1|$ and one obtains
\begin{equation}
P_{\mu \nu}(k_1,k_2) \; := \; 
g^2 \int_{-\infty}^\infty \frac{dp_1}{4\pi} \;
\frac{2-2\cos(p_1)}{p_1^2} \frac{2\cosh(p_1)-2}{p_1^2} e^{ip_1k_1}
e^{-|p_1|k_2} \frac{1}{|p_1|} \widehat{R}_{\mu \nu}(p_1,-i|p_1|) \; .
\end{equation}
Evaluating $\widehat{R}_{\mu \nu}(p_1,-i|p_1|)$ gives the matrix
\begin{equation}
\widehat{R}(p_1,-i|p_1|) \; = \; 4 \theta(p_1) 
\left[ \begin{array}{cc}
\; (e^{p_1} - 1)^2 \; &
\; \; \; -(e^{ip_1} - 1)(e^{p_1} -1) \; \\
\; -(e^{-ip_1} - 1)(e^{p_1} -1)  \; &
\; \; \; (e^{ip_1} - 1)(e^{-ip_1} - 1)  \;
\end{array} \right] \; ,
\end{equation}
where $\theta$ denotes the Heavyside step-function. The resulting
matrix is 
hermitean and thus can be diagonalized by some unitary matrix $U$. 
One finds
\begin{equation}
\widehat{R}(p_1,-i|p_1|) \; = \; 4 \theta(p_1) \;
U \; \mbox{diag}
\Big((e^{p_1} \! - \! 1)^2 \! + \! 2 \! - \! 2\cos(p_1)\; , \; 0 \Big) \;
U^\dagger \; ,
\end{equation}
where
\begin{equation}
U \; := \; [ (e^{p_1} \! - \! 1)^2 \! + \! 2 \! - \! 2\cos(p_1)
]^{-\frac{1}{2}}
\left[ \begin{array}{cc}
\; e^{p_1} - 1  \; &
\; e^{ip_1} - 1 \; \\
\; 1 - e^{-ip_1} \; &
\; e^{p_1} - 1 \; \end{array}
\right] \; .
\end{equation}
As can be seen from the explicit diagonalization (3.23), 
the matrix $\widehat{R}(p_1,-i|p_1|)$ 
has a vanishing and a 
positive real eigenvalue. The semidefiniteness of this matrix 
allows to construct a positive measure $d\mu[\lambda]$ as needed 
for the last term in (3.4). Putting things together
\begin{equation}
W^{mix}[A] \; = \; 
g^2 \int_{0}^\infty \frac{dp_1}{\pi} \;
\frac{2-2\cos(p_1)}{p_1^2} \frac{2\cosh(p_1)-2}{p_1^2} \;
\Big[ \Theta f(p_1) \Big] f(p_1) \; ,
\end{equation}
where we defined
\begin{equation}
f(p_1) \; = \; 
\sum_{n_1 \in Z\hspace{-1.35mm}Z} \sum_{n_2 > 1}
e^{-ip_1 n_1} e^{-p_1 n_2} \Big[
\frac{e^{p_1} - 1}{p_1} A_1(n_1,n_2) - 
\frac{e^{ip_1} - 1}{p_1} A_2(n_1,n_2) \Big] \; .
\end{equation}
Together with (3.6), Equation (3.25) establishes the 
decomposition of $-W^{lat}[A]$ into the form (3.4) and thus OS positivity
(3.3). OS positivity is the starting point of
the Osterwalder-Schrader reconstruction \cite{osterschra} which 
leads to Hamiltonian and physical Hilbert space.

It is remarkable that the proof given does not restrict the values 
of the Jackiw-Rajaraman parameter $a$. This surprising result might be 
understood, by the fact that $a$ enters the continuum action 
(2.9), (2.10) for the gauge field only via a contact term, which is irrelevant
for the Osterwalder-Schrader reconstruction in the continuum.

%
%

\section{Critical points and continuum limit}
\setcounter{equation}{0}
In the case of the vectorlike Schwinger model, it turned out, that 
for the analysis of critical points it is most convenient to study
the two point function of the field strength \cite{latsch, stara}. Here 
we follow this strategy and compute
\begin{equation}
E(n) \; := \; \Big\langle F_{12}(n) F_{12}(0) \Big\rangle \; .
\end{equation}
For notational convenience we introduce new 
fields defined as 
\begin{equation}
\varphi_1(n) \; := \; A_2(n+\hat{e}_1) - A_2(n) 
\; \; , \; \; 
\varphi_2(n) \; := \; A_1(n+\hat{e}_2) - A_1(n) \; ,
\end{equation}
which implies (compare (2.2))
\begin{equation}
F_{12}(n) \; = \; \varphi_1(n) \; - \; \varphi_2(n) \; .
\end{equation}
In momentum space one finds
\begin{equation}
\widehat{\varphi}_1(p) \; = \; (e^{ip_1} - 1) \widehat{A}_2(p) 
\; \; \; \; , \; \; \; \; 
\widehat{\varphi}_2(p) \; = \; (e^{ip_2} - 1) \widehat{A}_1(p) \; .
\end{equation}
From now on we restrict ourself to the non-compact formulation 
(2.1) of the gauge field action. Rewriting it first to momentum space
and then in terms of $\varphi_\mu$ one obtains for $S_{nc}$
\begin{equation}
S_{nc} \; = \; \frac{1}{2} \int_{-\pi}^{\pi} \frac{d^2q}{(2\pi)^2} \; 
\widehat{F}_{12}(-q) \widehat{F}_{12}(q) \; = \; 
\frac{1}{2} \int_{-\pi}^{\pi} \frac{d^2q}{(2\pi)^2} \; 
\widehat{\varphi}_{\mu}(-q) \; P_{\mu \nu} \; \widehat{\varphi}_{\nu}(q)
\; ,
\end{equation}
where $P_{\mu \nu}$ is defined as $P_{\mu \nu} := 2\delta_{\mu \nu} -1$.
Inserting (4.4) in (2.11) and adding the result to (4.5), one obtains 
the effective action $S^{eff}_{nc}$ (see (3.6)) in terms 
of $\varphi_\mu$
\begin{equation}
S^{eff}_{nc} \; = \; 
\frac{1}{2} \int_{-\pi}^{\pi} \frac{d^2q}{(2\pi)^2} \; 
\widehat{\varphi}_{\mu}(-q) \; \widehat{C}^{-1}_{\mu \nu}(q) \; 
\widehat{\varphi}_{\nu}(q) \; ,
\end{equation}
where
\begin{eqnarray}
\widehat{C}^{-1}_{11}(q) \; & := & \; 1 + 
g^2 \Big(2-2\cos(q_1)\Big) \Big(2-2\cos(q_2)\Big)
\sum_{k \in Z\hspace{-1.35mm}Z^2}  
\frac{1}{r_1^4 \; r_2^4 \; r^2}
\Big[(a+1)r_2^2 r^2 - 2r_2^4 + i2r_1r_2^3 \Big] \; , \nonumber \\
\widehat{C}^{-1}_{12}(q) \; & := & \; - 1 + 
g^2 \Big(2-2\cos(q_1)\Big) \Big(2-2\cos(q_2)\Big)
\sum_{k \in Z\hspace{-1.35mm}Z^2}  
\frac{1}{r_1^4 \; r_2^4 \; r^2}
\Big[-2r_1^2 r_2^2 + i( r_1^3r_2 - r_1r_2^3) \Big] \; , \nonumber \\
\widehat{C}^{-1}_{21}(q) \; & := & \; 
\widehat{C}^{-1}_{12}(q) \; , \nonumber \\
\widehat{C}^{-1}_{22}(q) \; & := & \; 1 + 
g^2 \Big(2-2\cos(q_1)\Big) \Big(2-2\cos(q_2)\Big)
\sum_{k \in Z\hspace{-1.35mm}Z^2 }  
\frac{1}{r_1^4 \; r_2^4 \; r^2}
\Big[(a+1)r_1^2 r^2 - 2r_1^4 - i2r_1^3r_2 \Big] \; . \nonumber \\
\end{eqnarray}
We introduced the abbreviation
\begin{equation}
r_\mu \; := \; q_\mu \; + \; 2\pi \; k_\mu \; ,
\end{equation}
where $k$ runs over $\mbox{Z\hspace{-1.35mm}Z}^2$, 
i.e. over all Brillouin zones.
This sum over all Brillouin zones comes from rewriting the momentum
integral over $\mbox{I\hspace{-0.7mm}R}^2$ 
which shows up in (2.11) as a sum of integrals 
over the first Brillouin zone only (note that $\widehat{A}_\mu(q)$,
and thus also $\widehat{\varphi}_\mu(q)$ are periodic with respect 
to shifting the argument to another Brillouin zone). 

We quantize the model by performing the path integral over the fields
$\varphi_\mu$. This gives rise to a Gaussian integral with 
covariance $C$ (see below). As in the continuum model (compare e.g. 
\cite{jara}), no gauge fixing term has to be added, since the 
non-gauge-invariant action (2.11) already fixes the gauge.
In order to compute the covariance, we have to analyze
for which values of $a$
$\widehat{C}^{-1}(q)$ can be inverted as a matrix.  Define
\begin{equation}
D(q) \; := \; \mbox{det} \; \widehat{C}^{-1} (q) \; =: \; 
g^2 (a-1) D_1(q) \; + \; g^4 a^2 D_2(q) \; .
\end{equation}
The terms $D_j(q), j =1,2$ are easy to compute but rather lengthy and
we quote them in the Appendix. There it is also shown (Lemma 1)
that at least for 
$a \geq 2$ the real part of $D(q)$ is strictly positive for 
all values of $q$ and for $g > 0$. This establishes
that $\widehat{C}^{-1}(q)$ can be inverted as a matrix.
In the Appendix it is shown furthermore, that this result
implies the existence of the propagator also for the original fields 
$A_\mu$ for $a \geq 2$. It has to be noticed that for the model 
in the continuum the determinant is real and positive for $a >1$. 
We were not able to extend the proof given in the Appendix to 
values $a>1$. However in the discussion of the continuum limit 
(see below), it will be shown that this difference in the range 
of $a$ is irrelevant. 

The propagator $\widehat{C}(q)$ is given by
\begin{equation}
\widehat{C}(q) \; = \; \frac{1}{D(q)} \; \left[
\begin{array}{cc}
\; \widehat{C}^{-1}_{22}(q) \; &
\; -\widehat{C}^{-1}_{12}(q) \; \\ & \\
\; -\widehat{C}^{-1}_{12}(q) \; &
\; \widehat{C}^{-1}_{11}(q) \;
\end{array} \right] \; .
\end{equation}
Using (4.3) one finds 
\[
E(n) \; = \;
\Big\langle \Big[ \varphi_1(n) - \varphi_2(n) \Big] 
\Big[ \varphi_1(0) - \varphi_2(0) \Big] \Big\rangle \; = \;
\int_{-\pi}^{\pi} \frac{d^2q \; e^{iqn}}{(2\pi)^2} \; \Big[
\widehat{C}_{11}(q) + \widehat{C}_{22}(q) - 2\widehat{C}_{12}(q) \Big]
\]
\[
= \; \int_{-\pi}^{\pi} \frac{d^2q\; e^{iqn}}{(2\pi)^2} \; 
\frac{g^2 (a\!-\!1)}{D(q)}
\sum_{k \in Z\hspace{-1.35mm}Z^2} \frac{2-2\cos(q_1)}{r_1^2}
\frac{2-2\cos(q_2)}{r_2^2} \frac{r^2}{r_1^2 \; r_2^2} 
\]
\begin{equation}
= \; \int_{-\pi}^{\pi} \frac{d^2q \; e^{iqn}}{(2\pi)^2} \; 
\frac{g^2(a\!-\!1)D_1(q)}{g^2 (a\!-\!1) D_1(q) \; + \; g^4 a^2 D_2(q)} \; .
\end{equation}
In the last step we used the definition of $D_1(q)$ (Equation(A.2)) and 
inserted (4.9) for $D(q)$. Rewriting the integrand one obtains 
\begin{equation}
E(n) \; = \; \int_{-\pi}^{\pi} \frac{d^2q \; e^{iqn}}{(2\pi)^2} 
\left[ 1 - m^2 \frac{1}{\sigma(q) + m^2} \right] \; = \; 
\delta_{n_1,0} \delta_{n_2,0} \; - \; 
m^2 \int_{-\pi}^{\pi} \frac{d^2q}{(2\pi)^2} 
\frac{e^{iqn}}{\sigma(q) + m^2} \; ,
\end{equation}
where we defined
\begin{equation}
m^2 \; := \; g^2 \frac{a^2}{a-1} \; ,
\end{equation}
and
\begin{equation}
\sigma(q) \; := \; \frac{D_1(q)}{D_2(q)} \; .
\end{equation}
In order to extract the exponential falloff of $E(n)$ which defines
the correlation length, the pole structure of the integrand in the last 
term of (4.12) 
has to be analyzed. As can be seen from (A.2), (A.3) the $D_j(q), j=1,2$
involve cosine terms. Thus for the exact computation of the poles one 
has to solve transcendental equations, which cannot be done in 
closed form. As in the case of the vectorlike Schwinger model 
(compare \cite{latsch, stara}) in the 
hybrid approach one is reduced to a perturbative analysis for small
$m^2$. In lowest order one has to compute the zeros of $\sigma(q)$ in
the first Brillouin zone ($BZ_1$).
From its definition (4.14) it is clear that $\sigma(p)$ becomes zero
either if $D_1(q)$ vanishes, or if $D_2(q)$ approaches $\infty$.
Inspecting (A.2) one immediately finds that $D_1(q)$ has no zeros in
$BZ_1$. From (A.3) it can be seen that 
$D_2(q)$ becomes infinite only for $q \rightarrow 0$.
However for $q \rightarrow 0$, $D_1(q) \rightarrow \infty$ as well, and
one has to count powers. Using the infrared behaviour of $D_j(q), j =1,2$
(Lemma 2 in the Appendix), one finds
\begin{equation}
\sigma(q) \; = \; \frac{q^2}{q_1^2 \; q_2^2}\Big( 1 + O(q^2) \Big) 
\left[ \frac{1}{q_1^2 \; q_2^2}\Big( 1 + O(q^2) \Big) \right]^{-1} 
\; = \; q^2 \; \Big( 1 + O(q^2) \Big) \; .
\end{equation}
Thus $\sigma(q)$ has only one zero in $BZ_1$ (at $q = 0$), 
and Equation (4.15) displays the 
behaviour of $\sigma(q)$ around it. For small $m$, the integrand in (4.12) 
has a pole at 
\begin{equation}
q^2 \; = \; - m^2 \; \Big( 1 + O(m^2) \Big) \; .
\end{equation}
This behaviour suggests that for small $m$ the model has a correlation 
length 
\begin{equation}
\xi \; = \; m^{-1} \; = \; \frac{1}{g} \; 
\frac{\sqrt{a-1}}{a} \; .
\end{equation}
$\xi$ becomes infinite for $g \rightarrow 0$ and arbitrary values
of $a \geq 2$ (note that we already excluded values $a < 2$).
Of course the perturbative analysis given above does not prove that
(4.17) holds. However there is a more elegant way, since for this simple
model the continuum limit can be controlled analytically. 
For the  moment we assume that $\xi$ is given 
by (4.17) and show that this assumption gives the correct continuum limit. 

We define our length scale $L_0$ to be proportional to the correlation length,
i.e. $L_0 := \lambda \xi$. A physical distance $|x|$
is measured in units of $L_0$ giving rise to $x := n/L_0$. The continuum
gauge coupling $g_c$ ($g_c = e_c/\sqrt{2\pi}$),
which has the dimension of a mass is defined as 
$g_c a_c/\sqrt{a_c - 1} := L_0 g a/\sqrt{a - 1}$. 
We also included a Jackiw-Rajaraman 
parameter $a_c$ for the continuum to match the continuum result completely.
Putting things together, we obtain for the ratio $n/\xi$
\begin{equation}
\frac{n}{\xi} \; \; = \; \; ng \; \frac{a}{\sqrt{a-1}} \; \; = \; \; n \; m 
\; \; = \; \; \mbox{const} \; \; = \; \; 
x g \; \frac{a_c}{\sqrt{a_c-1}} \; \; =: \; \; x \; m_c \; .
\end{equation}

We now perform the continuum limit by sending $\xi \rightarrow \infty$,
($\Leftrightarrow g \rightarrow 0$) keeping $n/\xi$ fixed, i.e. we take the 
joint limit $g \rightarrow 0, n \rightarrow \infty$ in the 
sense of (4.18). We will show that 
this limit reproduces the continuum result for the two point function 
of the field strength. The continuum result reads (it can be computed 
easily using (2.10))
\begin{equation}
\Big\langle F_{12}(x) F_{12}(0) \Big\rangle_{cont} \; = \; 
\delta^{(2)}(x) \; - \; \frac{m^2_c}{2\pi} 
\mbox{K}_0(m_c |x|) \; .
\end{equation}
We define
\begin{equation}
I(n,g) \; = \; \int_{-\pi}^{\pi} \frac{d^2q}{(2\pi)^2} \; 
\frac{e^{iqn}}{\sigma(q) + m^2} \; \; \; \; \; , \; \; \; \; \; 
J(n,g) \; = \; \int_{-\pi}^{\pi} \frac{d^2q}{(2\pi)^2} \;
\frac{e^{iqn}}{q^2 + m^2} \; .
\end{equation}
$I(n,g)$ is just the integral in the last term of the expression for the 
two point function of $F_{12}$ on the lattice. The main step in 
controlling the continuum limit is to show that $d(n,g)$ defined as
\begin{equation}
d(n,g) \; := \; I(n,g) - J(n,g) \; = \; 
\int_{-\pi}^{\pi} \frac{d^2q\; e^{iqn}}{(2\pi)^2}
\frac{q^2 - \sigma(q)}{\Big( \sigma(q) + m^2 \Big)\Big( q^2 + m^2 \Big)} 
\; ,
\end{equation}
vanishes in the joint limit $g \rightarrow 0, n \rightarrow \infty$ 
taken in the sense of (4.18). Since $\sigma(q)$ becomes zero only
at $q = 0$, $d(n,0)$ exists (use (4.15) to see that the infrared singularity
cancels), and
\begin{equation}
\lim_{n \rightarrow \infty} d(n,0) \; = \; 0 \; ,
\end{equation}
due to the Riemann-Lebesgue lemma. Using the triangle inequality
\begin{equation}
\Big|d(n,g)\Big| \; \leq \; \Big| d(n,g) - d(n,0) \Big| + 
\Big| d(n,0) \Big| \; ,
\end{equation}
the problem is reduced to showing that $\Big| d(n,g) - d(n,0) \Big| 
\rightarrow 0$ when performing the joint limit 
$g \rightarrow 0, n \rightarrow \infty$.
\begin{equation}
\Big| d(n,g) - d(n,0) \Big| \; \leq \; m^2
\int_{-\pi}^{\pi} \frac{d^2q}{(2\pi)^2} \left|
\frac{1 - q^2/\sigma(q)}{q^2} \right| \; 
\left| \frac{q^2 + \sigma(q) + m^2}{ 
\Big( \sigma(q) + m^2 \Big)\Big( q^2 + m^2 \Big) } \right| \; .
\end{equation}
The first factor in the integrand could become unbounded 
only for $q \rightarrow 0$ or 
$\sigma(q) \rightarrow 0$. For the first case ($q \rightarrow 0$) 
$\sigma(q)$ behaves as $q^2(1+O(q))$ (see (4.15)),
and thus the factor is bounded for 
$q \rightarrow 0$. This already covers the second case, since 
$\sigma(q)$ has its only zero at $q = 0$, as discussed above. 
Thus there exists a bound $A$ for the first factor. We obtain
\[
\Big| d(n,g) - d(n,0) \Big| \; \leq \; m^2 A
\int_{-\pi}^{\pi} \frac{d^2q}{(2\pi)^2} \; 
\left| \frac{q^2 + \sigma(q) + m^2}{ 
\Big( \sigma(q) + m^2 \Big) \Big( q^2 + m^2 \Big) } \right|
\]
\begin{equation}
= \; m^2 A
\int_{-\pi}^{\pi} \frac{d^2q}{(2\pi)^2} \; 
\frac{1}{ q^2 + m^2} \; + \; m^2 A
\int_{-\pi}^{\pi} \frac{d^2q}{(2\pi)^2} \; 
\frac{1}{ q^2 + m^2} \; \left| \frac{q^2}{\sigma(q) + m^2} \right| \; .
\end{equation}
The second factor in the integrand of the last integral in (4.25) 
also can be bounded using the properties of $\sigma(q)$.
\begin{equation}
\left| \frac{q^2}{\sigma(q) + m^2} \right| \; = \; 
\left| \frac{q^2/\sigma(q)}{1 + m^2/\sigma(q)}\right| \; = \;
\left| \frac{q^2}{\sigma(q)} \right| \frac{1}{\sqrt{
1 + 2 \mbox{Re}\;  m^2/\sigma(q) + m^4/|\sigma(q)|^2 }} \; .
\end{equation}
The first factor $|q^2/\sigma(q)|$ 
can be seen to be bounded, since $\sigma(q)$ 
becomes zero only at $q=0$ and behaves as $q^2(1+O(q))$ there (see (4.15)).
Thus there exists some bound $B$ with $|q^2/\sigma(q)| \leq B$ for 
$q \in BZ_1$. In the Appendix it is shown that $D_2(q)$
has a positive real part (see Lemma 1). Using the fact that $D_1(q)$ is real 
and positive (see (A.2)) and the definition (4.14) of $\sigma(q)$, one 
concludes that $m^2/\sigma(q)$ has a positive real part. Thus the 
argument of the square root in the last term in (4.26) is greater 1.
The last factor in (4.26) is bounded by 1, and the whole term by 
$B$. Putting things together, one obtains
\[  
\Big| d(n,g) - d(n,0) \Big| \; \leq \; m^2 A (1+B)
\int_{-\pi}^{\pi} \frac{d^2q}{(2\pi)^2} \; 
\frac{1}{ q^2 + m^2} 
\]
\begin{equation}
\leq \; \frac{m^2 A (1+B)}{(2\pi)^2} \int_0^{2\pi} d\varphi
\int_0^{2\pi} dr \frac{r}{r^2 + m^2} \; = \; 
\frac{m^2 A (1+B)}{4\pi} \Big[\ln(4\pi^2 + m^2) - \ln(m^2) \Big] \; 
\stackrel{g \rightarrow 0}{\longrightarrow} \; 0 \; .
\end{equation} 
We used the fact that the integrand is positive, and extended the 
area of integration to a circle of radius $2\pi$ around the origin. 
Inserting (4.22) and (4.27) into (4.23) establishes that 
$|d(n,g)| \rightarrow 0$ or equivalently $I(n,g) \rightarrow J(n,g)$ when 
performing the joint limit $g \rightarrow 0, n \rightarrow \infty$ 
in the sense of (4.18). Finally, using the transformation 
$p := q/m$ of the integration variable, $J(n,g)$ can 
be rewritten to (use (4.18))
\begin{equation}
J(n,g) \; = \; 
\int_{-\pi/m}^{\pi/m} \frac{d^2p}{(2\pi)^2} \; 
\frac{e^{ipxm_c}}{ p^2 + 1} \; = \; \frac{1}{2\pi} 
\mbox{K}_0(m_c |x|) \; + \; O(g) \; .
\end{equation}
We conclude
\begin{equation}
I(n,g) \; \longrightarrow \; \frac{1}{2\pi} 
\mbox{K}_0(m_c |x|) \; ,
\end{equation}
when taking the joint limit $g \rightarrow 0, n \rightarrow \infty$.
The last step is to remove the extra factor $m^2$ which shows up in
(4.12) in front of $I(n,g)$ with a wave function renormalization constant
\begin{equation}
Z_{12}(g) \; := \; \frac{m_c^2}{m^2} \; .
\end{equation}
This also reproduces the contact term of the continuum result (4.19),
since $\delta_{n_1,0} \delta_{n_2, 0} m_c^2/m^2 \rightarrow 
\delta^{(2)}(x)$ for $g \rightarrow 0, n \rightarrow \infty$ with 
$n m = x m_c$ held fixed. Thus we have established
\begin{equation}
\lim_{g \rightarrow 0, n \rightarrow \infty} \;
Z_{12}(g) \Big\langle F_{12}(n) F_{12}(0) \Big\rangle\; = \; 
\Big\langle F_{12}(x) F_{12}(0) \Big\rangle_{cont} \; .
\end{equation}
It has to be remarked that the result (4.31) for the continuum limit
is an exact result, and contains no more perturbation expansion. 
It shows, that the model has critical points 
where the continuum limit can be performed at $g=0, a \geq 2$.

%
%

\section{Remarks on fermionic observables}
\setcounter{equation}{0}
Since the fermions are treated in the continuum, the continuum
fermion propagator in an external interpolated field has to be 
computed. The continuum propagator $G(x,y;A)$ has to obey the 
Greens function equation
\begin{equation}
\left[\frac{\partial}{\partial x_1} - i \frac{\partial}{\partial x_2}
- ie \Big(A_1(x) - i A_2(x) \Big) \right] \; G(x,y;A) 
\; = \; \delta^{(2)}(x-y) \; .
\end{equation}
A solution can be found easily by using Schwinger's original ansatz
\cite{schwinger}
\begin{equation}
G(x,y;A) \; := \; G_0(x-y) \; 
\exp\Big( ie [\Phi(x) - \Phi(y)] \Big) \; ,
\end{equation}
where $G_0$ denotes the propagator for free, left handed fermions 
which obeys 
\begin{equation}
[ \partial_1 - i\partial_2 ] G_0(x) \; = 
\; \delta^{(2)}(x-y) \; ,
\end{equation}
and $\Phi(x)$ is a solution of
\begin{equation}
[ \partial_1 - i\partial_2 ] \Phi_0(x) \; = 
\; A_1(x) - i A_2(x) \; .
\end{equation}
Using $[ \partial_1 - i\partial_2 ][ \partial_1 + i\partial_2 ]
= \triangle$ and the Greens function $C(x)$ of $-\triangle$ given 
by $C(x) = -\ln(\mu^2 x^2)/4\pi$ ($\mu$ parametrizes the freedom in the 
infrared regularization) one obtains
\begin{equation}
G_0(x) \; = \; - [ \partial_1 + i\partial_2 ] C(x) \; = \; 
\frac{1}{2\pi} [x_1 - ix_2]^{-1} \; ,
\end{equation}
and 
\[
\Phi(x) = - \int_{-\infty}^\infty d^2 z \; C(x-z) 
\Big[\frac{\partial}{\partial z_1} + i \frac{\partial}{\partial z_2}
\Big] \; \Big[ A_1(z) - i A_2(z) \Big] 
\]
\begin{equation} 
= \; 
- \int_{-\infty}^\infty \frac{d^2p}{(2\pi)^2} \; 
\frac{e^{ipx}}{p^2} [ ip_1 - p_2 ] \; [\widetilde{A}_1(p) - i 
\widetilde{A}_2(p)] \; .
\end{equation}
Inserting the interpolated gauge field $\widetilde{A}_\mu^{int}$ (2.6)
and restricting the arguments $x,y$ to lattice points $n,m$, one ends up 
with
\begin{equation}
G(n,m;A^{int}) \; = \; \frac{1}{2\pi} 
\Big[ (n_1-m_1) - i(n_2-m_2) \Big]^{-1} \; 
\exp\Big( ie [\Phi^{int}(n) - \Phi^{int}(m)] \Big) \; ,
\end{equation}
where
\[
\Phi^{int}(n) \; := \; \int_{-\pi}^{\pi} 
\frac{d^2q \; e^{iqn}}{(2\pi)^2} \Bigg[ 
\widehat{A}_1(q) \sum_{k \in Z\hspace{-1.35mm}Z}
\frac{2-2\cos(q_2)}{r_2^2} \frac{e^{-iq_1}-1}{r_1} \frac{r_1 + ir_2}{r^2} 
\]
\begin{equation}
- \; i \widehat{A}_2(q) \sum_{k \in Z\hspace{-1.35mm}Z}
\frac{2-2\cos(q_1)}{r_1^2} \frac{e^{-iq_2}-1}{r_2} \frac{r_1 + ir_2}{r^2}
\Bigg] \; .
\end{equation}
Again the integral over $\mbox{I\hspace{-0.7mm}R}^2$ had to be rewritten 
as an integral 
over the first Brillouin zone only, since the lattice fields and not the
continuum fields 
are integrated over in the path integral. 

The basic, gauge invariant fermionic observable is
\begin{equation}
\chi(n) \; U_{\cal C}(n,m) \; \overline{\chi(m)} \; ,
\end{equation}
where $U_{\cal C}$ denotes a gauge transporter along some contour
${\cal C}$. It has to be remarked that one cannot build $U_{\cal C}$
out of the lattice gauge transporters (2.4). The reason is that 
such a simple construct ($U_{\cal C}^{naive}$) 
transforms under a lattice gauge transformation
\begin{equation}
A_\mu(n) \; \longrightarrow \; A_\mu(n) \; + \;  
\Lambda(n+\hat{e}_\mu) - \Lambda(n) \; ,
\end{equation}
as
\begin{equation}
U_{\cal C}^{naive}(n,m) \; \; \longrightarrow \; \;
e^{-ie\Lambda(n)} \; \; U_{\cal C}^{naive} \; \; e^{ie\Lambda(m)} \; ,
\end{equation}
while the propagator $G(n,m;A^{int})$ transforms as 
\begin{equation}
G(n,m;A^{int}) \; \longrightarrow \; 
e^{ie\Delta(n)} \; G(n,m;A^{int}) \; e^{-ie\Delta(m)} \; ,
\end{equation}
with
\begin{equation}
\Delta(n) \; := \; \int_{-\pi}^{\pi} 
\frac{d^2q \; e^{iqn}}{(2\pi)^2} 
\frac{2-2\cos(q_1)}{q_1^2} 
\frac{2-2\cos(q_2)}{q_2^2} 
\widehat{\Lambda}(q) \; .
\end{equation}
Thus the naive lattice transporter does not make (5.9) gauge invariant
under lattice gauge transformations. Instead one has to use
the continuum gauge transporter along some contour on the lattice
and evaluate it for the interpolated fields $A^{int}_\mu$ 
\begin{equation}
U_{\cal C}(n,m) \; = \; \exp\left(
ie\int_{\cal C} A^{int}_\mu(z) dz_\mu \right) \; := \; 
\exp\left(ie (A^{int}_\mu, j_\mu^{\cal C} ) \right) \; .
\end{equation}
For e.g. a rectangular contour ${\cal R}$ with length $2s$ and height $s$ 
($s \in \mbox{I\hspace{-0.74mm}N}$) one 
obtains 
\begin{eqnarray}
\widehat{j^{\cal R}_1}(q) \; & := & \; - i 2 \sum_{k \in Z\hspace{-1.35mm}Z}
\frac{2-2\cos(q_2)}{r_2^2} \; \frac{e^{iq_1}-1}{r_1} \; 
\frac{ \sin(q_1 s) }{r_1} \; e^{-iq_2s} \; ,\nonumber \\
\widehat{j^{\cal R}_2}(q) \; & := & \; - i 2 \sum_{k \in Z\hspace{-1.35mm}Z}
\frac{2-2\cos(q_1)}{r_1^2} \; \frac{e^{iq_2}-1}{r_2} \; 
\sin(q_1 s) \; \frac{1-e^{-iq_2s}}{r_2} \; .
\end{eqnarray}
Thus for computing expectation values of products of (5.9), only
Gaussian functional integrals have to be solved. However the
results become rather involved, since in the covariance (4.10) as well
as in the exponentials of propagator (5.7) and gauge transporter (5.14)
sums over all Brillouin zones occur. 

We conclude this section with computing at least one fermionic
expectation value, namely one where the integration over the gauge
fields becomes trivial. We consider the two point function of the chiral 
density 
\begin{equation}
\Big\langle \chi(n) \overline{\chi}(n) \; 
\chi(0) \overline{\chi}(0) \Big\rangle^c \; ,
\end{equation}
where the superscript $c$ stands for connected.
One finds 
\begin{equation}
\Big\langle \chi(n) \overline{\chi}(n) \; 
\chi(0) \overline{\chi}(0) \Big\rangle^c \; = \; 
- \int \frac{D[A]}{Z} \; G(0,n;A^{int}) \; G(n,0;A^{int}) \; .
\end{equation}
$D[A]/Z$ denotes a normalized functional integral over the lattice 
gauge fields, which can e.g. be realized as a Gaussian integral over 
$\varphi_\mu$ with covariance (4.10), as was done in the last section. 
However, by inspecting (5.7), one finds that the dependence on the 
gauge fields cancels for the quoted product of propagators and 
integration over the gauge fields simply gives a factor 1. We end up with
\begin{equation}
\Big\langle \chi(n) \overline{\chi}(n) \; 
\chi(0) \overline{\chi}(0) \Big\rangle^c \; \; = \; \;
- \; \frac{1}{(2\pi)^2} [n_1 -i n_2]^{-2} \; .
\end{equation}
As can be seen from this result, the two point function of the 
chiral density clusters as in the continuum model. No condensate is 
being formed. This behaviour is 
different from the vectorlike Schwinger model where the expectation 
value of the chiral density is known to be nonvanishing, in the continuum 
as well as in hybrid approach \cite{latsch}.

From (4.18) and (5.18) it can be seen that again just a wave function 
renormalization is necessary in the continuum limit.
Define
\begin{equation}
Z_{\chi \overline{\chi}} (g) \; := \; \frac{m_c^2}{m^2} \; .
\end{equation}
One ends up with
\begin{equation} 
\lim_{g \rightarrow 0, n \rightarrow \infty} \;
Z_{\chi \overline{\chi}} (g) \;
\Big\langle \chi(n) \overline{\chi}(n) \;
\chi(0) \overline{\chi}(0) \Big\rangle^c \; = \; 
\Big\langle \chi(x) \overline{\chi}(x) \;
\chi(0) \overline{\chi}(0) \Big\rangle^c_{cont} \; \; = \; \; 
- \; \frac{1}{(2\pi)^2} [x_1 -i x_2]^{-2} \; ,
\end{equation}
where the limit was taken in the sense of (4.18)
%
%

\section{Concluding remarks}
\setcounter{equation}{0}
It has been demonstrated that the hybrid approach works rather well 
for the chiral Schwinger model. The effective lattice gauge theory was 
constructed using the Jackiw-Rajaraman determinant for the chiral fermions
in a background field. The resulting effective lattice gauge theory was 
proven to be Osterwalder-Schrader positive, for both the non-compact 
and the Wilson formulation. For the non-compact formulation we established 
the existence of critical points for zero gauge coupling and 
Jackiw-Rajaraman parameter $a \geq 2$. The continuum limit was performed
explicitly for the two point functions of field strength and 
chiral densities. 

As a next step it might be interesting to explore the hybrid approach for
a model with non-Abelian gauge group. A possible candidate could be 
QCD$_2$ where reasonable explicit definitions of regularized continuum 
fermion determinants in a background field exist (for an overview see e.g.
\cite{abdalla}). Even more challenging is of course the extension to four 
dimensional theories. The problem there is that the determinant 
is either given very implicitly or only the first few terms of a loop
expansion are known. An approach that is promising at least 
for vectorlike theories is the introduction of two independent 
lattice cutoffs, a coarse lattice
for the gauge fields and a finer lattice for the fermions \cite{hernan}.
It should be possible to proof OS positivity for this setting by 
conventional methods \cite{oster}.
\vskip3mm
\noindent
{\bf Ackknowledgement :} \\
The author thanks Erhard Seiler for his ongoing interest in this project
and valuable discussions, and Peter Weisz for reading the manuscript. 

%
%
\newpage
\appendix
\renewcommand{\thesection}{Appendix}
\section{}
\renewcommand{\thesection}{A}
\setcounter{equation}{0}
In the Appendix we analyze the properties of the determinant of the kernel 
$\widehat{C}^{-1}(q)$ of the effective action (4.6). 
The determinant of the matrix $\widehat{C}^{-1}(q)$ can be written as 
\begin{equation}
D(q) \; := \; \mbox{det} \widehat{C}^{-1} (q) \; := \;  
g^2 (a+1) D_1(q) \; + \; g^4 a^2 D_2(q)  \; .
\end{equation}
The terms $D_j(q), j=1,2$ are given by
\begin{equation}
D_1(q) \; = \; \sum_{k \in Z\mbox{\hspace{-1.35mm}}Z^2} 
\frac{2-2\cos(r_1)}{r_1^2}
\frac{2-2\cos(r_2)}{r_2^2} \frac{r^2}{r_1^2 \; r_2^2} \; , 
\end{equation}
and 
\[
D_2(q) \; := \; \frac{1}{a^2} \sum_{k,l \in Z\hspace{-1.35mm}Z}
\frac{\Big( 2 - 2\cos(q_1) \Big)^2}{r_1^2 \; s_1^2} \; 
\frac{\Big( 2 - 2\cos(q_2) \Big)^2}{r_2^2 \; s_2^2} \;
\frac{1}{r_1^2 \; r_2^2 \; r^2} \; \frac{1}{s_1^2 \; s_2^2 \; s^2} 
\]
\[
\times \Big\{ \Big[(a+1) r_2^2 r^2 - 2 r_2^4 \Big]
\Big[(a+1) s_1^2 s^2 - 2 s_1^4 \Big]
+ 4 r_1 s_1^3 r_2^3 s_2 - 4 r_1^2 s_1^2 r_2^2 s_2^2
+ (r_1^3 r_2 - r_1 r_2^3) (s_1^3 s_2 - s_1 s_2^3) 
\]
\begin{equation}
+ i 2\Big[
r_1 r_2^3 \Big( (a+1) s_1^2 s^2 - 2 s_1^4 \Big) -
s_1^3 s_2 \Big( (a+1) r_2^2 r^2 - 2 r_2^4 \Big) +
r_1^2 r_2^2 (s_1^3 s_2 - s_1 s_2^3) +
s_1^2 s_2^2 (r_1^3 r_2 - r_1 r_2^3) \Big] \Big\} \; .
\end{equation}
We introduced the abbreviations
\begin{equation}
r_\mu \; := \; q_\mu + 2\pi k_\mu \; \; \; \; , \; \; \; \; 
s_\mu \; := \; q_\mu + 2\pi l_\mu \; .
\end{equation}
\vskip3mm
\noindent
{\bf Lemma 1: } 
{\sl The real part of $D(q)$ is strictly positive for $a \geq 2$ and
$g > 0$.}
\vskip3mm
\noindent
From (A.2) it is obvious, that $D_1(q)$ is real and strictly positive. Thus
there is left to show, that $D_2(q)$ has a positive real part.
The real part of $D_2(q)$ is the sum over 
\begin{equation}
S \; := \; \Big[(a+1) r_2^2 r^2 - 2 r_2^4 \Big]
\Big[(a+1) s_1^2 s^2 - 2 s_1^4 \Big]
+ 4 r_1 s_1^3 r_2^3 s_2 - 4 r_1^2 s_1^2 r_2^2 s_2^2
+ (r_1^3 r_2 - r_1 r_2^3) (s_1^3 s_2 - s_1 s_2^3) \; ,
\end{equation}
with the positive weight
\begin{equation}
\frac{\Big( 2 - 2\cos(q_1) \Big)^2}{r_1^2 \; s_1^2} \; 
\frac{\Big( 2 - 2\cos(q_2) \Big)^2}{r_2^2 \; s_2^2} \;
\frac{1}{r_1^2 \; r_2^2 \; r^2} \; \frac{1}{s_1^2 \; s_2^2 \; s^2} \; .
\end{equation}
Note that the weight is symmetric with respect to $r$ and $s$. 
This allows to interchange the role of $r$ and $s$ in individual terms
of $S$ under the sum. Using this trick one can rewrite $S$ to 
\begin{equation}
S \; = \; [(a-1)^2 - 1] r^2 \; s^2 \; r_1^2 \; s_2^2 \; + \; 
2(a-1) r_1^2 \; r_2^2 \; (s^2)^2 \; + \; \frac{1}{2} r^2 \; s^2 \;
(r_1 s_2 + r_2 s_1)^2 \; .
\end{equation}
Equation (A.7) makes it obvious, that each term in $D_2(q)$ has
a positive real part for $a \geq 2$, and thus Lemma 1 is proven.

Lemma 1 implies that the determinant $D(q)$ does not vanish for 
$a \geq 2$, and that the propagator $C(q)$ for the fields 
$\varphi_\mu$ exists for that range of $a$. It also implies, that 
the propagator for the original gauge fields exists. From (4.4) it
can be seen, that the determinant of the kernel for the action written
in terms of $A_\mu$ differs from $D(q)$ by an overall factor 
$[2-2\cos(q_1)][2-2\cos(q_2)]$. This factor is positive and vanishes
in the first Brillouin zone only at $q=0$. The corresponding zero is
cancelled in $D_1(q)$ and $D_2(q)$ in at least the $k=0$ terms,
which is sufficient to obtain a determinant with strictly positive
real part. Thus the kernel in the action written in terms of the gauge 
fields can be inverted for $a \geq 2$ as well, giving the 
propagator for the gauge field $A_\mu$.
\vskip3mm
\noindent
{\bf Lemma 2: } 
{\sl The terms $D_j(q), j=1,2$ have the following infrared behaviour:}
\begin{equation}
D_1(q) \; = \; \frac{q^2}{q_1^2 \; q_2^2} \Big[1 + O(q^2)\Big]
\; \; \; \; , \; \; \; \;
D_2(q) \; = \; \frac{1}{q_1^2 \; q_2^2} \Big[1 + O(q^2)\Big] \; .
\end{equation}
\vskip3mm
\noindent
From the definition (A.2) it is clear that the most singular term
in the sum for $D_1(q)$ is the term with $k = 0$. It behaves as 
$q^2/( q_1^2 q_2^2 )$. All other terms have at least an extra factor
$q_\mu^2$ which already establishes the first part of (A.8). 
The infrared behaviour of $D_2(q)$ is a little bit more involved to 
analyze. $D_2(q)$ can be rewritten as the sum of 
\begin{equation}
\sum_{k,l \in Z\hspace{-1.35mm}Z}
\frac{\Big( 2 - 2\cos(q_1) \Big)^2}{r_1^2 \; s_1^2} \; 
\frac{\Big( 2 - 2\cos(q_2) \Big)^2}{r_2^2 \; s_2^2} \;
\frac{1}{r_1^2 \; s_2^2} \; ,
\end{equation}
and
\[
\frac{1}{a^2} \sum_{k,l \in Z\hspace{-1.35mm}Z}
\frac{\Big( 2 - 2\cos(q_1) \Big)^2}{r_1^2 \; s_1^2} \; 
\frac{\Big( 2 - 2\cos(q_2) \Big)^2}{r_2^2 \; s_2^2} \;
\frac{1}{r_1^2 \; r_2^2 \; r^2} \; \frac{1}{s_1^2 \; s_2^2 \; s^2} 
\]
\[
\times \Big\{2a \; r_1^2 s_2^2 ( s_1^2 r_2^2 - r_1^2 s_2^2 ) 
- r_1^2 r_2^2 s_2^4 + r_1 s_1 r_2^3 s_2^3 
- r_1^4 s_1^2 s_2^2 + r_1^3 s_1^3 r_2 s_2
+ r_1^4 s_2^4 -3 r_1^2 s_1^2 r_2^2 s_2^2 
\]
\begin{equation}
+ 2 r_1^3 s_1 r_2 s_2^3
+i2(a-1) \Big[r_1^4 s_1 s_2^3 - r_1^3 s_1^2 r_2 s_2^2
+ r_1^2 s_1 r_2^2 s_2^3 - r_1^3 r_2 s_2^4 \Big] \Big\} \; .
\end{equation} 
The first part (A.9) is easily seen to have its most singular term at 
$k=l=0$ and it behaves as $1/( q_1^2 q_2^2 )$. The second term (A.10)
is more subtle.
For $k=l=0$, the contribution adds up to zero, and the potentially
most singular term is not there. The terms $k_\mu \neq 0;
l_\mu \neq 0; k_1,k_2 \neq 0; l_1,l_2 \neq 0$ and $k_\mu, l_\nu \neq 0$
can be seen to be of $O(q^2)$ less singular than 
$1/( q_1^2 q_2^2 )$. This establishes
the second part of (A.8).

%
%

\end{document}